\documentclass[twocolumn]{aastex631}

\usepackage{graphicx}
\usepackage{float}

\usepackage{subfigure}

\usepackage     [utf8]                  {inputenc}
\usepackage     [T1]                    {fontenc}
\usepackage{chngcntr}
\usepackage                             {graphicx}
\usepackage     [english]               {babel}
\usepackage                             {outlines}
\usepackage{xcolor}

\begin{document}

\title{The Evolution of Half-Mass Radii and Color Gradients for Young and Old Quiescent Galaxies at $0.5 < z < 3$ with JWST/PRIMER}

\author[0009-0001-4005-5490]{Maike Clausen}
\affiliation{Institute for Astronomy, University of Edinburgh, Royal Observatory, Blackford Hill, Edinburgh, EH9 3HJ, UK}
\affiliation{Max Planck Institute of Astronomy, Königstuhl 17, 69117 Heidelberg, Germany}

\author[0000-0003-1665-2073]{Ivelina Momcheva}
\affiliation{Max Planck Institute of Astronomy, Königstuhl 17, 69117 Heidelberg,Germany}

\author[0000-0001-7160-3632]{Katherine E. Whitaker}
\affiliation{Department of Astronomy, University of Massachusetts, Amherst, MA 01003, USA}
\affiliation{Cosmic Dawn Center (DAWN), Denmark} 

\author[0000-0002-7031-2865]{Sam E. Cutler}
\affiliation{Department of Astronomy, University of Massachusetts, Amherst, MA 01003, USA}

\author[0000-0001-5063-8254]{Rachel S. Bezanson}
\affiliation{Department of Physics and Astronomy and PITT PACC, University of Pittsburgh, Pittsburgh, PA, USA}

\author[0000-0002-1404-5950]{James S. Dunlop}
\affiliation{Institute for Astronomy, University of Edinburgh, Royal Observatory, Blackford Hill, Edinburgh, EH9 3HJ, UK}

\author[0000-0001-9440-8872]{Norman A. Grogin}
\affiliation{Space Telescope Science Institute, 3700 San Martin Drive, Baltimore, MD 21218, USA}

\author[0000-0002-6610-2048]{Anton M. Koekemoer}
\affiliation{Space Telescope Science Institute, 3700 San Martin Drive, Baltimore, MD 21218, USA}

\author[0000-0003-4368-3326]{Derek McLeod}
\affiliation{Institute for Astronomy, University of Edinburgh, Royal Observatory, Blackford Hill, Edinburgh, EH9 3HJ, UK}

\author{Ross McLure}
\affiliation{Institute for Astronomy, University of Edinburgh, Royal Observatory, Blackford Hill, Edinburgh,EH9 3HJ, UK}

\author[0000-0001-8367-6265]{Tim B. Miller}
\affiliation{Center for Interdisciplinary Exploration and Research in Astrophysics (CIERA), Northwestern University, 1800 Sherman Ave, Evanston, IL 60201, USA}

\author[0000-0002-7524-374X]{Erica Nelson}
\affiliation{Department for Astrophysical and Planetary Science, University of Colorado, Boulder, CO 80309, USA}

\author[0000-0002-5027-0135]{Arjen van der Wel}
\affiliation{Ghent University, Krijgslaan 281, Building S9, 9000 Ghent, Belgium}

\author[0000-0002-6047-1010]{David Wake}
\affiliation{Department of Physics and Astronomy, University of North Carolina Asheville, 1 University Heights, Asheville, NC 28804, USA}

\author[0000-0003-3735-1931]{Stijn Wuyts}
\affiliation{Department of Physics, University of Bath, Claverton Down, Bath BA2 7AY, UK}

\begin{abstract}
We present a study of the size growth of the red sequence between $0.5<z<3,$ tracing the evolution of quiescent galaxies in both effective half-light and half-mass radii using multi-wavelength JWST/NIRCam imaging provided by the PRIMER survey.
Half-light radii are measured from imaging in 6 different filters for 455 quiescent galaxies with log($M_*/M_{\odot}$)$>10$, whereas half-mass radii are derived from the F444W profiles together with the F277W-F444W color-$M_*$/L relation.
We investigate the dependence of the ratio $r_{e, \mathrm{mass}}/r_{e, \mathrm{light}}$ on redshift, stellar mass, and the wavelength used to measure $r_{e, \mathrm{light}}$, also separating the sample into younger and older quiescent galaxies.
Our data demonstrate that rest-frame infrared sizes accurately trace mass-weighted sizes while sizes measured at rest-frame optical wavelengths (0.5-0.7$\mu$m) are 0.1-0.2 dex larger, with only minor variations in redshift.  
We find that the average size of young quiescent galaxies agrees with that of old quiescent galaxies at intermediate masses, $10<$log($M_*/M_{\odot}$)$<11$, within their respective uncertainties in all observed-frame half-light, rest-frame half-light and half-mass radius measurements.
At face value, our results point to a combination of progenitor bias and minor mergers driving the size growth of intermediate-mass quiescent galaxies at $0.5<z<3$.
Our results further indicate that the varying contributions to the general quiescent population by young and old quiescent galaxies can mimic evolution in redshift.

\end{abstract}

\section{Introduction}
The morphologies of quiescent galaxies are closely linked to the processes that shape them. 
For this reason, structural parameters like the half-light radius can be used to infer information about the mechanisms driving their evolution. 
Numerous studies demonstrate that the average half-light radius of quiescent galaxies increases towards later cosmic time \citep[e.g.,][]{cimatti2008, vandokkum2008, damjanov2009, vandersande2013, patel2017, mowla2019, mathaduru2020,clausen20243ddash}.
This growth of the red sequence has been attributed to either progenitor bias: recently quenched galaxies at lower redshift are larger because the star-forming galaxies out of which they form are more extended compared to the star-forming population at earlier cosmic time \citep[e.g.,][]{vandokkum2008, vanderwel2009, szomoru2011, Carollo2013,poggianti2013}, or to the combination of minor mergers and the continuous accretion of stellar mass \citep[e.g.,][]{Trujillo2006b, bezanson2009, naab2009, vandokkum2010, Patel2013, hill2016, newman12, belli15, suess23}.
Comparing the sizes of recently quenched galaxies to those quenched a long time ago provides crucial insight into which processes dominate the evolution of the quiescent population as a whole. 
Because star-forming galaxies are, on average, larger than quiescent galaxies \citep[e.g.,][]{szomoru2011, VanderWel2014, Barro2017}, progenitor bias predicts young quiescent galaxies also to be larger than their older counterparts \citep[e.g.,][]{Ichiwaka2017, pawlik2019, Wu2020}. 
Alternatively, assuming minor mergers are a significant contributor to the growth of the red sequence, we would expect young quiescent galaxies to be smaller than their older counterparts \citep[e.g.,][]{bezanson2009, vandokkum2010, newman12, suess23}.
We demonstrate, in \cite{clausen20243ddash}, that young quiescent galaxies appear to be, on average, smaller than old quiescent galaxies at $1.5<z<3$ \citep[see also][]{almaini2017,chen2022}.
We also demonstrate that this size difference is most pronounced in the most massive sub-sample (log($M_*/M_{\odot}$) $>$ 11) and diminishes at intermediate stellar masses. 
At face value, these results favor a scenario in which massive quiescent galaxies quench by a brief galaxy-wide burst of star-formation and continue to grow through minor mergers and accretion \citep[e.g.,][]{newman12, Ceverino2015, wellons2015}.

At the same time, studies have shown that light profiles tend to exhibit larger effective radii than mass profiles \citep[e.g.,][]{guo2011,szomoru2011,fang2013,suess2019a, Suess2022, miller2023} which can bias conclusions based on light profile measurements.
The trends in half-light radii are dependent on the radial distribution of stellar populations, dust, and metallicity and can impact the light profiles across a range of wavelengths \citep[e.g.,][]{wu2005,labarbera2009, wuyts2010, Tortora2011, Conroy2017, sanchez2019, vandokkum2021}.
Since these effects produce similar changes in both the color and $M_*/L$ gradients \citep[e.g.,][]{Rix1993,bell_deJong2001,chabrier2003IMF, Portinari2004}, color-$M_*/L$ relations can be used to convert light profiles into mass profiles.
Several studies \citep[e.g.,][]{szomoru2013, fang2013, zibetti2009,meidt2014} use two-dimensional surface brightness fitting in multi-wavelength imaging to extract rest-frame color profiles. 
They employ spectral energy distribution (SED) fitting to determine an empirical rest-frame color-$M_*/L$ relation that results in radial $M_*/L$ profiles. 
These are then converted into radial mass profiles, which are integrated to determine the half-mass radius.
However, this approach assumes a universal distribution of age, star-formation timescale, and dust extinction across the galaxy, introducing systemic uncertainties as some quenching processes might cause radial variations in stellar populations.
To account for this, \cite{Mosleh2017} and \cite{suess2019a} follow the same approach but perform SED fitting on multiple concentric elliptical annuli, allowing for radial variations in these parameters.
To improve simple color-$M/_*L$ relations, \cite{vanderwel2024} use light profiles from different filters to derive a multi-color-$M_*/L$ relation. 
\cite{miller2023} also develop an alternative method of extracting radial color gradients by fitting a cascade of Gaussians to the measurements of surface brightness profiles (SBPs) in two filters, improving on parameterized approaches that would fit, e.g., S\'{e}rsic profiles \citep[][]{sersic1963} to the measured SBPs. 

In this work, we attempt to shed light on the growth of quiescent galaxies by deriving light- and mass-weighted sizes using high-resolution multi-wavelength JWST/NIRCam imaging from the PRIMER survey. We select a sample of 508 quiescent galaxies utilizing the photometry described in Section \ref{sec:half_mass_sample}. 
Following a method similar to that of \cite{szomoru2013} and \cite{miller2023}, we derive half-mass radii in Section \ref{sec:half_mass_method}.
Section \ref{sec:half_mass_results} demonstrates the structural evolution of mass-weighted sizes, color gradients, and the dependences of these properties on total stellar mass, which we discuss in section \ref{sec:half-mass-discussion}.
Finally, we close in section \ref{sec:half-mass_conclusion} with a summary of the key results.

We assume a $\Lambda$CDM cosmology with $\Omega_M = 0.3$, $\Omega_{\Lambda}=0.7$ and $H_0 = 70$ km s$^{-1}$ Mpc$^{-1}$, as well as a \citet{chabrier2003IMF} initial mass function for stellar masses.
\section{Data and Sample Selection}\label{sec:half_mass_sample}
\begin{figure*}
    \centering
    \includegraphics[width=\textwidth]{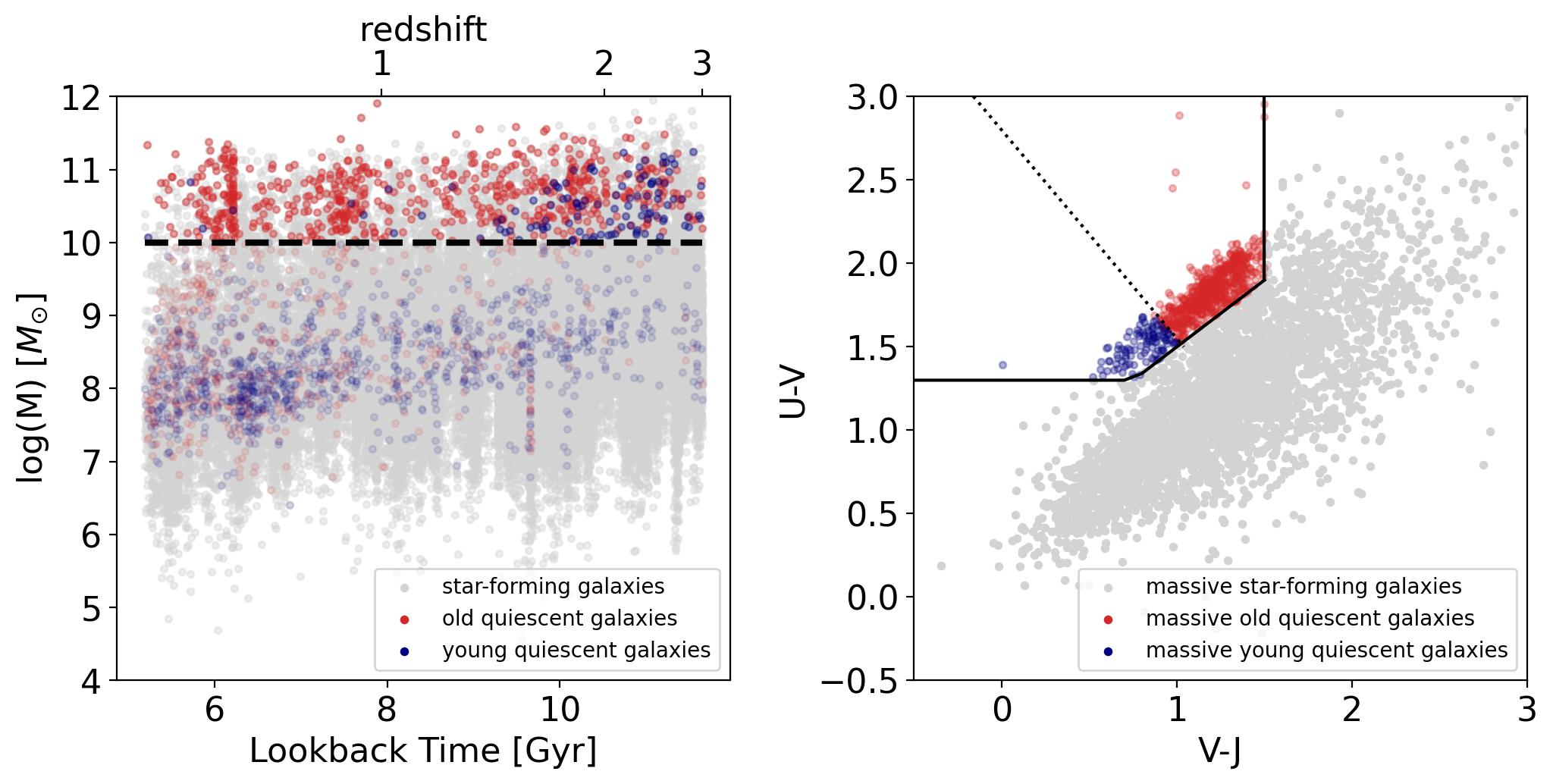}
    \caption[Evolution of mass of quiescent galaxies in the PRIMER survey]{\textbf{Left panel :} Star-forming (gray), old quiescent (red), and young quiescent (blue) galaxies in PRIMER. Since the most massive galaxies drive the rest-frame optical size difference, we select only galaxies that have masses higher than log$M_* > 10M_{\odot}$ (black dashed line). \textbf{Right panel:} Selection of quiescent galaxies (only massive log$M_* > 10M_{\odot}$ sample depicted) within (V-J)-(U-V) space (black line). The dotted line divides the sample into recently quenched (blue) and older (red) quiescent galaxies.}
    \label{fig:sample_half_mass}
\end{figure*}

Our sample selection is based on catalogs derived from the PRIMER survey ((JWST-GO-1837), PI: \citep{Dunlop2021primerproposal}). 
The JWST Cycle 1 general observing program PRIMER has obtained multi-wavelength JWST/NIRCam imaging in the F090W, F115W, F150W, F200W, F277W, F356W, and F444W filters in parts of the COSMOS and UDS fields covering a total area of 378 arcmin$^2$.
The sample is selected from a photometric catalog (Cutler et al., in prep., private communications) that was constructed using \texttt{APERPY} \citep[][]{Weaver2023aperpy} following the methodology of \citet{weaver2024}.
The redshifts and stellar masses included in the catalog were derived using the \texttt{PROSPECTOR-$\beta$} \citep[][]{Wang2023prospector}.
Using the provided flags securing reliable photometry and photometric redshifts, we restrict the selection to the redshift range of interest: $0.5<z<3$ and choose a mass limit of log($M_*$/$M_{\odot}$) $>$ 10 (dashed black line in Figure \ref{fig:sample_half_mass}) motivated by an observed bimodality in the stellar mass distribution within the quiescent galaxy population (see below).

The left panel of Figure \ref{fig:sample_half_mass} shows star-forming (grey), young (blue), and old (red) quiescent galaxies in the PRIMER survey between $0.5<z<3$.
There is a distinct scarcity of quiescent galaxies at 9 $\lesssim$ log($M_*$/$M_{\odot}$) $\lesssim$ 10, causing a bimodality in the mass distribution.
Though the majority of lower mass galaxies are faint and have therefore a small signal-to-noise ratio, the depicted galaxies were selected with SNR$>$5.
However, the gradual buildup of the stellar mass function of low-mass quiescent galaxies has previously suggested the possibility of discrete populations of quiescent galaxies.
In contrast, high-mass passive galaxies have been observed to be fully quenched at high redshift \citep[e.g.,][]{peng2010,ilbert2013, Davidzon2017, santini2022, weaver2023, Cutler2024}.
The bimodality is thought to be intrinsic to the quiescent population and not an artifact of the data set \citep[see, e.g.,][]{Cutler2024}.
As this study focuses on the more massive galaxies, the bimodality motivates the limitation of the galaxies in this sample to be above log($M_*$/$M_{\odot}$) $>$ 10.

We use rest-frame UVJ colors from the catalog to select quiescent galaxies and divide the sample into young and old quiescent galaxies based on their stellar ages inferred from their position in the UVJ diagram (see Figure \ref{fig:sample_half_mass}) \citep[e.g.,][]{whitaker2012, whitaker2013, belli2019,akhshik2023, clausen20243ddash}.

We perform a two-dimensional morphological analysis using \texttt{GALFIT} \citep{peng2002} on cutouts from the image mosaics in all PRIMER bands to derive half-light radii, S\'{e}rsic indices, magnitudes, position angles, projected axis ratios, and central positions of the galaxies. 
A more detailed description of this is included in the next Section.
We exclude any galaxy that cannot be securely fit (35\%) in all six available filters.
The final sample contains 426 old and 39 young quiescent galaxies, which are used for the subsequent analysis in this chapter.
\section{Methodology}\label{sec:half_mass_method}
\begin{figure*}
    \centering
    \includegraphics[width=\textwidth]{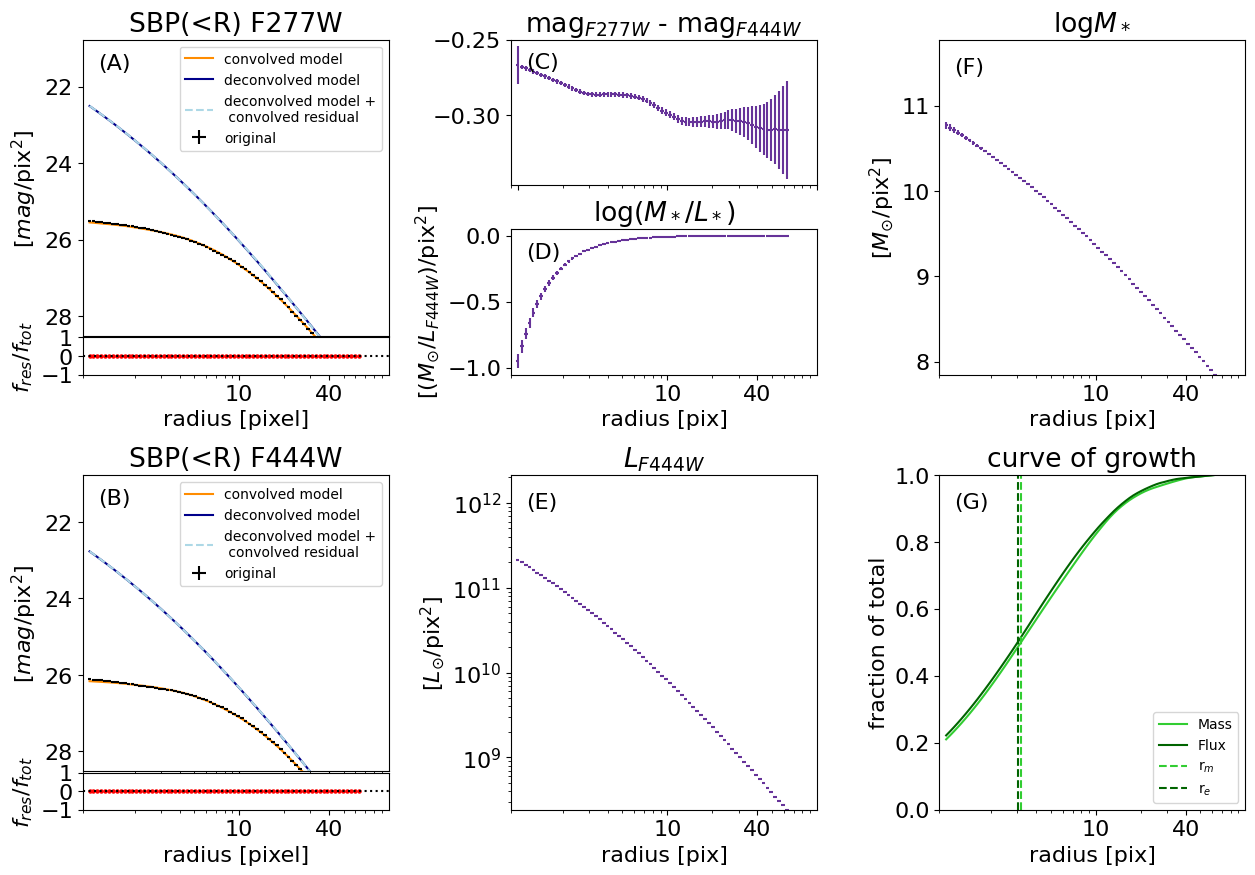}
    \caption[Methodology of deriving effective half-mass radii]{Methodology of deriving half-mass radii. \textbf{Panels (A) and (B):} Measured SBP (black), convolved model from \texttt{GALFIT} (orange) and deconvolved model (dark blue) in two photometric bands (F227W and F444W, respectively). The residual between the flux from the original cutout and the convolved model is added to the deconvolved model (bright blue). This is considered the total flux. The lower panels display in red the magnitude of the residual correction to be small compared to the total flux. \textbf{Panel (C):} Radial color profile derived from the total flux in F277W (A) and F444W (B). \textbf{Panel (D):} Derived radial $M_*$ / $L_{444}$ profile using the color-$M_*$ / $L_{444}$ relation derived in Subsection \ref{subsec:derive_ML}. \textbf{Panel (E):} Radial luminosity profile derived from the total flux measured in F444W. \textbf{Panel (F):} Radial mass profile derived from multiplying the radial $M_*$ / $L_{444}$ profile (D) with the luminosity profile (E). \textbf{Panel (G):} Curve of growth for mass (bright green) and total flux in F444W  (dark green) with half-light and half-mass radii (dashed lines). The uncertainties of the half-mass radii are determined with Monte-Carlo simulations.}
    \label{fig:half_mass_method}
\end{figure*}

To derive half-mass radii, we follow an approach similar to that of \citet{szomoru2011} \citep[see also][]{fang2013, Chan2016, Mosleh2017, vanderwel2024}.
This method \citep[also corresponding to the third method explored in][]{suess2019a} consists of the following steps:
First, we measure radial SBPs in several photometric filters converted into radial color profiles.
Next, we derive an empirical relation between the integrated colors of the galaxies and their $M_*/L$ ratios, which we use to convert the color profiles into radial $M_*/L$ profiles. 
Then, we multiply the radial $M_*/L$ profiles with the derived luminosity profiles to derive radial mass profiles.
Finally, we calculate the curve of growth for the radial mass profile and determine the half-mass radius as the radius containing half of the total mass.
These steps are demonstrated in Figure \ref{fig:half_mass_method} and described in more detail below.

\subsection{Deriving Surface Brightness Profiles}\label{subsec:derive_SBPs}
We create cutouts for each galaxy of 5$\times $$R_{kron}$ of the science image, the weight map, the exposure time map, and the segmentation map.
In the case of extremely compact galaxies in our sample, we set a lower limit of at least 125 pixels to sample the SBP of these galaxies securely.
The segmentation map is used to create a mask to prevent contamination by bad pixels or nearby bright objects.
PRIMER is a composite survey of multiple pointings leading to overlapping regions. 
To account for these varying exposure times, an error map is calculated by combining the sky background variance (1/$wht$) and Poisson noise ($sci/exp$):
\begin{equation}\label{eq:err_map}
    err = \sqrt{1/wht + sci/exp},
\end{equation}
where $wht$, $sci$, and $exp$ are cutouts of the weight, science, and exposure maps.
To accurately account for the point spread function (PSF) in each band, we use PSFs constructed from unresolved stars in the survey \citep[see, e.g.,][]{Cutler2024}.
We use \texttt{GALFIT} \citep{peng2002} to perform two-dimensional morphological fitting assuming a S\'{e}rsic profile.
Next, we follow \citet{szomoru2010} and perform a first-order correction on the models by measuring the residual flux from the convolved model in concentric apertures based on the ellipticity and central position from the \texttt{GALFIT} outputs.
Doing so allows for the correction of the S\'{e}rsic model where it deviates from the galaxy's true light profile.
The parameters for S\'{e}rsic index, magnitude, and half-light radius are used to reconstruct a deconvolved radial surface brightness profile to which the radial residual flux is added.

Panel (A) and (B) in Figure \ref{fig:half_mass_method} show the measured radial profiles of the original science image (black) and the convolved model (orange), which agree reasonably well.
The dark blue line displays the deconvolved one-dimensional S\'{e}rsic profile derived from the fit parameters to which the residual flux profile (measured from the two-dimensional residual between the science cutout and the convolved \texttt{GALFIT} model) is added (bright blue).
The errors are measured from the error map (Eq. \ref{eq:err_map}) using the same apertures to measure the residuals and observed profiles.
Following previous studies, we proceed using the residual corrected radial profiles to derive half-mass radii, mitigating the uncertainty of assuming a particular shape of the radial mass profile.
To do so, we use the residual corrected SBPs (panels (A) and (B) in Figure \ref{fig:half_mass_method}) to derive radial color profiles (panel (C) necessary to extract a radial $M_*/L$ profile (panel D).

\subsection{Determining the $M_*/L$-color relation}\label{subsec:derive_ML}
We follow the approach of \cite{miller2023} and \cite{vanderwel2024} and use observed colors by fitting the integrated color-$M_*/L$ relation in different bins of redshift.
To do so, we first calculate color-$M_*/L$ for F090W, F115W, F150W, F200W, F277W, and F356W with respect to F444W and find the tightest correlation with the F277W-F444W colors (spearman rank of -0.96 compared to lower correlation coefficients in the other filter combinations with the lowest being -0.85 in F356W-F444W colors).
Next, we split our sample into bins of redshift and perform a least-squares fit to extract the color-$M_*$/$L_{444}$ relation for each bin in the form of:
\begin{equation}\label{eq:color_relation}
log(M_*/L_{F444W}) = -2.5\times \log_{10}\frac{F(277W)}{F(444W)}\times A + C
\end{equation}
where A and C are the coefficients and constants for which we fit. 
\begin{figure*}
    \centering
    \includegraphics[width=\textwidth]{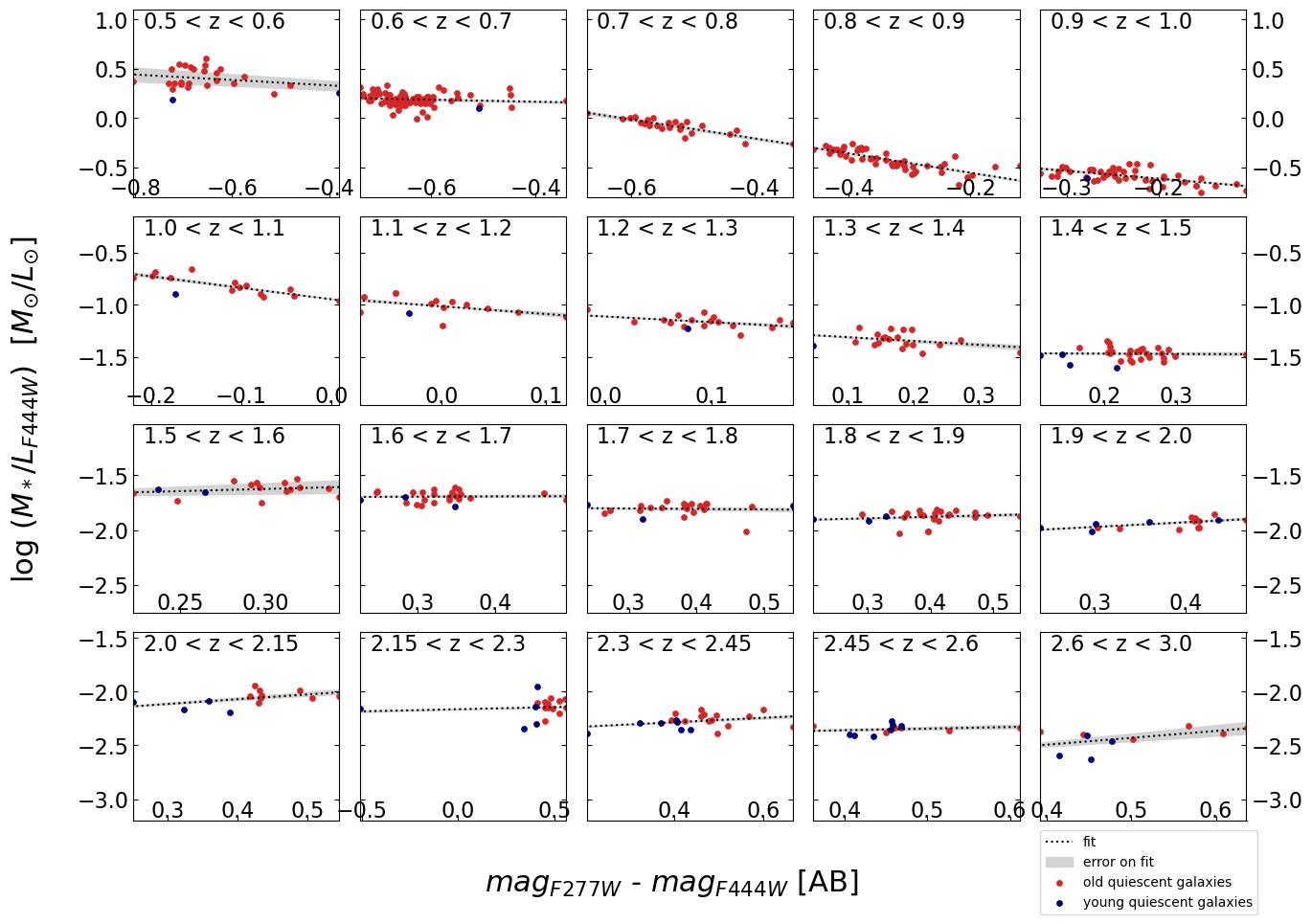}
    \caption{$M_*$ / $L_{444}$ dependence on redshift. Data points are color-coded by the classification of a galaxy as young quiescent (blue) or old quiescent (red). The fit parameters for each redshift bin are listed in Table \ref{tab:color_ML_fit}.}
    \label{fig:color_ML_277_444}
\end{figure*}
The detailed redshift binning allows for a decrease in scatter to 0.06 dex around the relation.
To ensure statistical accuracy, we make sure that every bin includes at least ten galaxies.
Figure \ref{fig:color_ML_277_444} demonstrates the fits whose fit parameters as a function of redshift can be found in Table \ref{tab:color_ML_fit}.

\begin{table}[h]\small
    \centering
    \begin{tabular}{c|c|c}
        redshift & A & C \\
        \hline
    0.5 $<$ z $<$ 0.6 & -0.2752 $\pm$ 0.0562 & 0.2215 $\pm$ 0.0246 \\
    0.6 $<$ z $<$ 0.7 & -0.1034 $\pm$ 0.009 & 0.1246 $\pm$ 0.0037 \\
    0.7 $<$ z $<$ 0.8 & -0.9578 $\pm$ 0.0153 & -0.5896 $\pm$ 0.0044 \\
    0.8 $<$ z $<$ 0.9 & -0.9817 $\pm$ 0.0135 & -0.7514 $\pm$ 0.0016 \\
    0.9 $<$ z $<$ 1.0 & -0.7895 $\pm$ 0.0259 & -0.7736 $\pm$ 0.0015 \\
    1.0 $<$ z $<$ 1.1 & -1.0943 $\pm$ 0.0606 & -0.946 $\pm$ 0.0011 \\
    1.1 $<$ z $<$ 1.2 & -0.7361 $\pm$ 0.1693 & -1.0138 $\pm$ 0.0005 \\
    1.2 $<$ z $<$ 1.3 & -0.5408 $\pm$ 0.0754 & -1.1118 $\pm$ 0.0008 \\
    1.3 $<$ z $<$ 1.4 & -0.365 $\pm$ 0.0651 & -1.2732 $\pm$ 0.0023 \\
    1.4 $<$ z $<$ 1.5 & -0.0274 $\pm$ 0.0441 & -1.4623 $\pm$ 0.0026 \\
    1.5 $<$ z $<$ 1.6 & 0.3952 $\pm$ 0.2258 & -1.7429 $\pm$ 0.0197 \\
    1.6 $<$ z $<$ 1.7 & 0.0233 $\pm$ 0.0311 & -1.7019 $\pm$ 0.0035 \\
    1.7 $<$ z $<$ 1.8 & -0.0437 $\pm$ 0.0356 & -1.789 $\pm$ 0.0051 \\
    1.8 $<$ z $<$ 1.9 & 0.1393 $\pm$ 0.0309 & -1.9337 $\pm$ 0.0049 \\
    1.9 $<$ z $<$ 2.0 & 0.4258 $\pm$ 0.0251 & -2.0991 $\pm$ 0.0037 \\
    2.0 $<$ z $<$ 2.15 & 0.444 $\pm$ 0.0576 & -2.2486 $\pm$ 0.0104 \\
    2.15 $<$ z $<$ 2.3 & 0.04 $\pm$ 0.0101 & -2.1637 $\pm$ 0.0022 \\
    2.3 $<$ z $<$ 2.45 & 0.2038 $\pm$ 0.0214 & -2.3661 $\pm$ 0.0045 \\
    2.45 $<$ z $<$ 2.6 & 0.1509 $\pm$ 0.0435 & -2.42 $\pm$ 0.0093 \\
    2.6 $<$ z $<$ 3.0 & 0.6406 $\pm$ 0.143 & -2.7505 $\pm$ 0.0358 \\      
    \end{tabular}
    \caption{Fit parameters for the color-$M_*$ / $L_{444}$ relation (Eq. \ref{eq:color_relation}) between F277W and F444W dependent on redshift}
    \label{tab:color_ML_fit}
\end{table}

We assume that these color-$M_*$ / $L_{444}$ relations are valid for integrated colors and radial trends inside each galaxy.
Therefore, we use the derived relations to extract radial $M_*$/$L_{444}$ profiles for the galaxies in our sample (see panels (C) and (D) in Figure \ref{fig:half_mass_method}).
The color errors are calculated using Gaussian error propagation. The errors in intrinsic $M_*$/$L_{444}$ profiles are a result of observing a scatter of 0.06 dex of the predicted integrated $M_*$ / $L_{444}$ ratios around the measured $M_*$/$L_{444}$ ratios.

\subsection{Derivation of Half-Mass Radii}\label{subsec:reivation_half-mass_radii}
To convert the radial $M_*$ / $L_{444}$ profiles into mass profiles, we first convert the F444W measured SBPs (Figure \ref{fig:half_mass_method}, panel B) into luminosity profiles (Figure \ref{fig:half_mass_method}, panel E) following Equation 24 in \cite{Hogg1999}:
\begin{equation}\label{eq:luminosity}
    L_{obs} = F_{obs} \times 4 \pi D_L^2 \times (1+z)^3.
\end{equation}
where $F_{obs}$ are the observed surface brightness measured in concentric anuli, $z$ is the photometric redshift from the catalogs and $D_L$ is the luminosity distance to redshift $z$ given the assumed cosmology.

The exponent of the redshift contribution results from a combination of a factor of $(1+z)^{-1}$ converting the emitted frequency to the observed frequency and the fact that the surface brightness of a redshifted object is reduced by $(1+z)^{4}$.
The errors are calculated using Gaussian error propagation, including the errors from the SBP measurements and assuming an error of 0.06 dex for the $M_*$ / $L_{444}$ profiles based on the scatter of the original measurement around the fitted color-$M_*$ / $L_{444}$ relations.

To extract the radial mass profile, we multiply the luminosity profile measured in F444W with the $M_*/L_{F444W}$ profiles.
This mass profile is then integrated to calculate a curve of growth from which the half-mass radius can be extracted.
We use Monte Carlo simulations with 1000 iterations to derive a distribution of possible half-mass radii for each galaxy whose 16th and 84th percentile are used as the uncertainties of the original measurement.
Figure \ref{fig:half_mass_method} demonstrates the different steps of the methodology.

\section{Results}\label{sec:half_mass_results}

\subsection{Evolution of Half-Mass Radii} \label{subsec:half-mass_radii_evolution}
Having derived half-mass and half-light radii for 455 quiescent galaxies in our sample, we first investigate the evolution of the average mass- and light-weighted sizes in the observed redshift range ($0.5<z<3$).
We calculate the geometric means of half-light and half-mass radii for old and young quiescent galaxies for five redshift bins. 
This ensures that at least ten galaxies per bin represent each population to warrant robust statistical analysis.
The half-light radii are measured in F150W imaging to allow reasonable comparison with previous studies measured at similar wavelengths.
\begin{figure*}
    \centering
    \includegraphics[width=\textwidth]{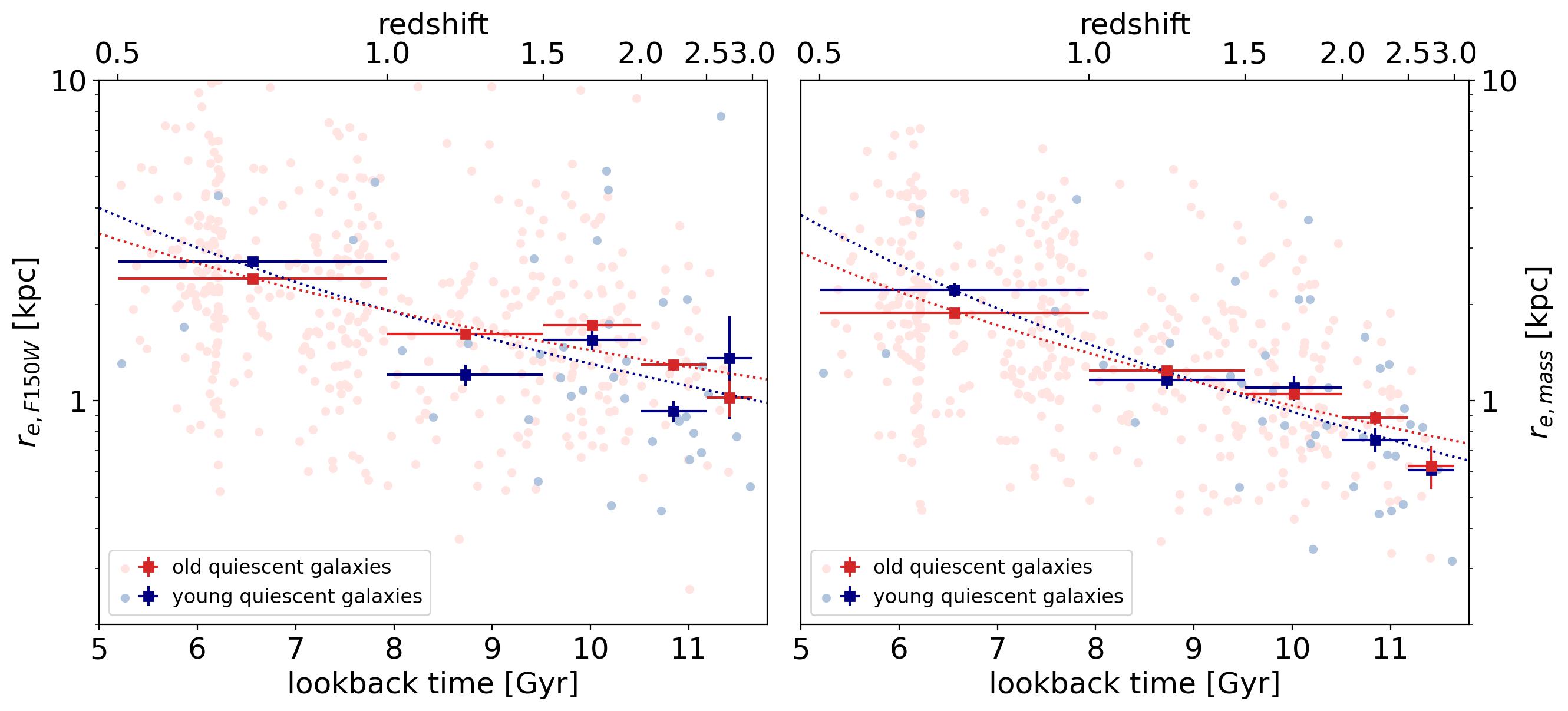}
    \caption{\textbf{Left panel}: Evolution of half-light radii measured at F150W of young (blue) and old (red) quiescent galaxies. \textbf{Right panel}: Evolution of half-mass radii. Both populations grow in both light-weighted and mass-weighted sizes towards lower redshift, with no significant difference. In both panels, the dotted lines depict the logarithmic fits to the respective data whereas the error bars depict the error in the mean (not the scatter).}
    \label{fig:half-mass_evolution}
\end{figure*}

The left panel of Figure \ref{fig:half-mass_evolution} shows a significant growth in average half-light radii towards lower redshift for both populations: 0.17 dex for the old quiescent sample and 0.23 dex for the young quiescent sample.
To quantify this growth we fit a logarithmic function $R \propto (1+z)^{\alpha}$ to both populations.
In agreement with \citep[][]{vanderwel2024}, we find exponents ($\alpha$) of (-1.85 $\pm$ -0.27) for young quiescent galaxies and (-1.35 $\pm$ 0.08) for old quiescent galaxies.
Considering the scatter of both populations and the small number of young quiescent galaxies, we regard the offset in size between young and old quiescent galaxies to be insignificant.
However, this is likely due to this sample being dominated by intermediate mass ($10<log(M_*/M_{\odot})<11$) galaxies, where the size difference is not as significant \citep[see][]{clausen20243ddash} and the fact that we have a small sample \citep[>10$\times$ smaller than][]{clausen20243ddash}.

The left panel of Figure \ref{fig:half-mass_evolution} demonstrates the previously observed growth in light-weighted sizes of the red sequence towards lower redshift, documented in the literature for two decades now \citep[e.g.,][]{Trujillo2004, Trujillo2006a, vanderwel2008, vandokkum2008, clausen20243ddash}.  
Here, we show that this growth is also prominent in mass-weighted size measurements for both of our quiescent samples. 
Our data shows an increase in the average mass-weighted size of 0.30 dex in the young quiescent compared to 0.35 dex in the old quiescent population between $0.5 < z < 3$.
We repeat the previous fit and find $R \propto (1+z)^{-2.33 \pm 0.05}$ for young quiescent galaxies and $R \propto (1+z)^{-1.81 \pm 0.04}$ for their older counterparts.
Both \cite{suess2019a} and \cite{miller2023} argue that correcting the $M/L$ gradients for redshift effects decreases the size evolution at $z>1$ when measured from optical colors.
In fact, \cite{miller2023} find no evolution of half-mass radius in the quiescent galaxy population while their half-light radii increase by 30\% between $1<z<2$.
In contrast, and in agreement with our measurements, \cite{vanderwel2024} observe substantial growth of the red sequence towards later cosmic times for light-weighted, mass-weighted, and deprojected sizes.
Since this growth of the red sequence has been widely observed in rest-frame optical size measurements (see references above), and now has been observed in half-mass radii as well \citep[this study,][]{mosleh2020, vanderwel2024}, it is likely to be a true evolution of the underlying population rather than an artifact of choosing light-weighted sizes in lieu of mass-weighted sizes as an observable.

Previous studies have suggested that the difference in half-light radius observed in some studies might be an effect of methodology rather than intrinsic morphology.
For instance, \cite{fang2013} and \cite{suess2019a} both suggest that light profiles exaggerate key structural differences and should be substituted with stellar mass profiles.
Later studies by \cite{Suess2022} used early NIRCam results to demonstrate that light-weighted size measurements taken at 4.4$\mu$m are up to 0.15 dex smaller than sizes measured at 1.5$\mu$m (we find a difference of 0.1 $\pm$ 0.01 dex in this study).
\cite{vanderwel2024} show that mass-weighted sizes are 0.1-0.15 dex smaller than light-weighted sizes.
We find a difference of 0.13$\pm$0.01 dex in this study.
Nonetheless, these studies look at the general quiescent galaxy population and do not distinguish between recently quenched and older subsamples.
It is possible that the color gradients causing the discrepancy between half-light and half-mass radius might differ.
If this were not the case, both samples would be similarly corrected according to their color gradients, leading to the size gap persisting in mass-weighted size measurements.
We, therefore, turn to the investigation of the evolution of the $r_{e, \mathrm{mass}}/r_{e, \mathrm{light}}$ ratio between the two populations.

\subsection{Rest-Frame Color Gradients}\label{subsec:colorgradient_restframe}

This sub-section studies the evolution of $r_{e, \mathrm{mass}}/r_{e, \mathrm{light}}$ ratios for both old and young quiescent galaxies. 
If the $r_{e, \mathrm{mass}}/r_{e, \mathrm{light}}$ ratio is smaller (or larger) than unity, the average half-mass radii are smaller (larger) than the respective average half-light radii indicating radially varying stellar populations.
For instance, a small $r_{e, \mathrm{mass}}/r_{e, \mathrm{light}}$ ratio suggests a substantial contribution from brighter stars at larger radii, indicating the presence of comparatively younger stars, in which case a strong color gradient would be present.
Color gradients are closely linked to $r_{e, \mathrm{mass}}/r_{e, \mathrm{light}}$ ratios.
Therefore, this section uses the two phrases interchangeably.

\begin{figure*}
    \centering
    \includegraphics[width=\textwidth]{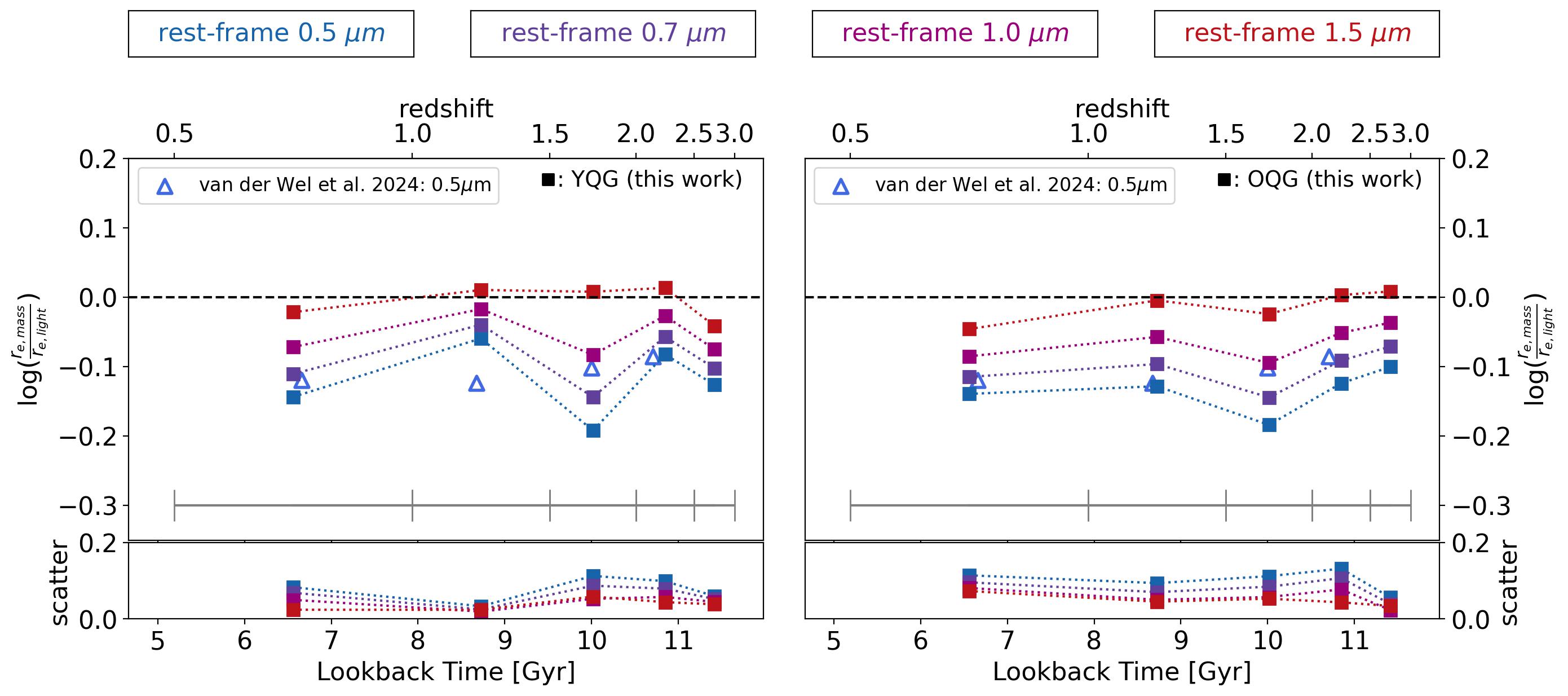}
    \caption{Color gradient evolution measured at rest-frame 0.5, 0.7, 1.0 and 1.5$\mu$m respectively. Both old (\textbf{right panels}) and young (\textbf{left panels}) quiescent galaxies show stronger, negative color gradients at bluer rest-frame wavelengths relative to measurements at longer wavelengths. Considering the scatter the two populations agree at all redshifts.}
    \label{fig:color-gradient_restrame}
\end{figure*}

For each galaxy, we fit the securely measured half-light radii for all available filters with rest-frame wavelengths larger than 4000\r{A} filter with a linear function and extrapolate the half-light radius at rest-frame wavelengths corresponding to 0.5$\mu$m, 0.7$\mu$m, 1.0$\mu$m, and 1.5$\mu$m.
Next, we divide the half-mass radii by the respective half-light radii measured at the four different rest-frame wavelengths and calculate the average ratio of $r_{e, \mathrm{mass}}/r_{e, \mathrm{light}}$ for five bins of redshift for each of the two populations of quiescent galaxies.
Figure \ref{fig:color-gradient_restrame} shows the evolution of color gradient for young quiescent (left panel) and old quiescent (right panel) galaxies with the extent of the bins demonstrated by the black bar at the bottom.
The data points are color-coded by the respective restframe wavelengths, with redder colors corresponding to measurements taken at longer wavelengths.

Figure \ref{fig:color-gradient_restrame} demonstrates stronger color gradients at shorter wavelengths (i.e., negative values) and increasingly neutral values at longer wavelengths (1.5$\mu$m) for both old and young quiescent galaxies.
Our data do not show a significant difference between old and young quiescent galaxies for the measured rest-frame  $r_{e, \mathrm{mass}}/r_{e, \mathrm{light}}$ ratios across the redshift range considered herein.
Considering the scatter in the distribution, and despite our lower mass limit ($M_* >10^{10} M_{\odot}$ compared to their $M_* >10^{11} M_{\odot}$) we find our measurements to agree with those taken from \citet{VanderWel2014}.

Old quiescent galaxies exhibit marginally stronger color gradients towards lower redshift (between 0.03 to 0.04 dex).
Though the scatter within the young quiescent population is of similar magnitude compared to the old quiescent galaxies, the small number of young quiescent galaxies combined with the observed jumps in $r_{e, \mathrm{mass}}/r_{e, \mathrm{light}}$ ratios preclude a robust determination of the trend towards neutral color gradients at higher redshift.
Reducing the number of redshift bins and repeating the analysis decreases the variety in means of color gradients and reveals the same tendencies as the older galaxies.
However, to conclusively determine the extent of this trend a much larger sample than the one at hand is needed.

\subsection{Stellar Mass Dependence}\label{subsec:half_mass_mass_dependence}
\begin{figure*}
    \centering
    \includegraphics[width=\textwidth]{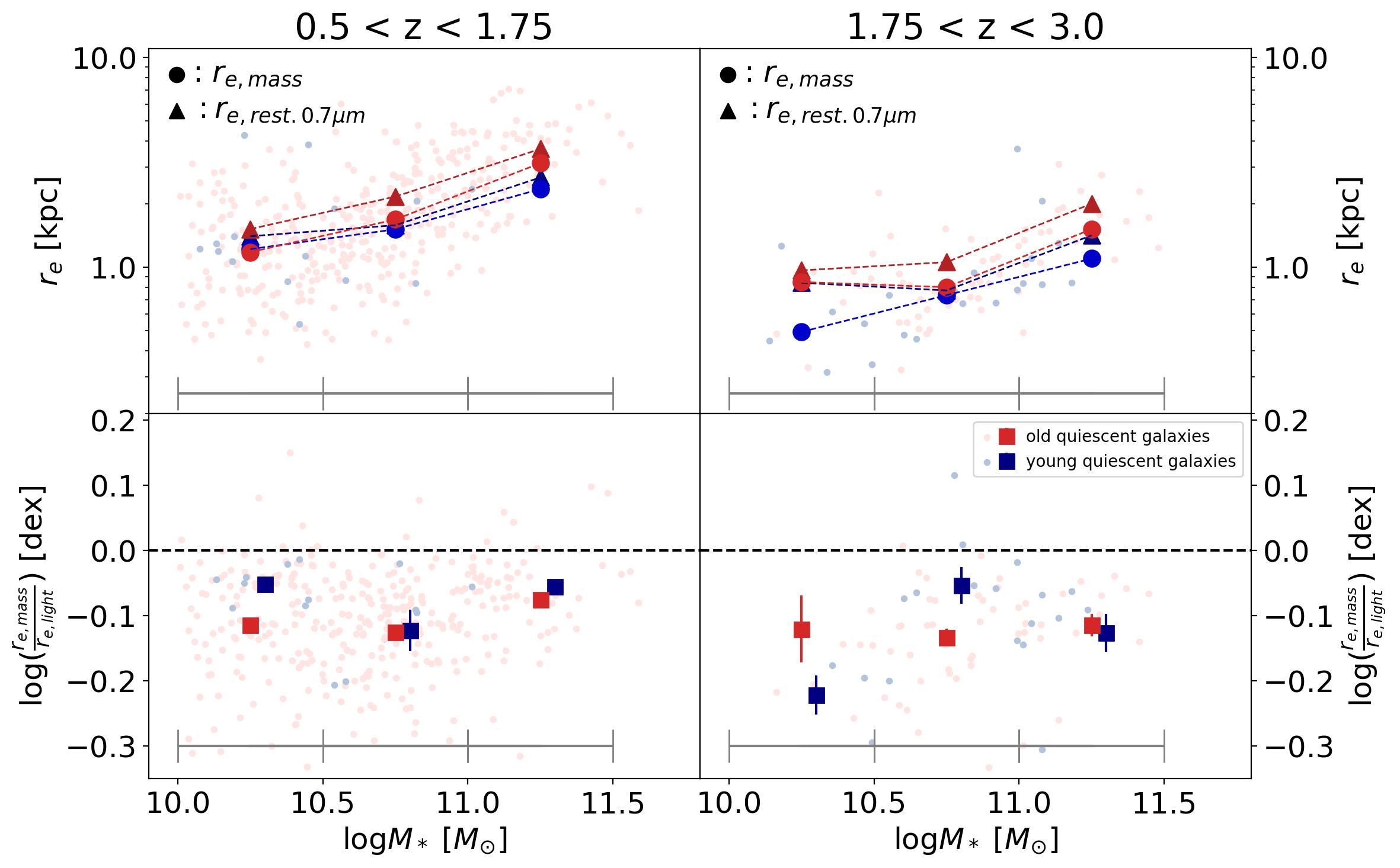}
    \caption[Dependence of half-light radius, half-mass radius and the ratio between the two on stellar mass and redshift]{The average half-mass and restframe 0.7$\mu$m half-light radii of both young and old quiescent galaxies increase with total stellar mass both at $0.5<z<1.75$ (\textbf{upper left panel}) and at $1.75<z<3$ (\textbf{upper right panel}). The $r_{e, \mathrm{mass}}/r_{e, \mathrm{light}}$ ratios of both populations agree within their uncertainties at all stellar masses and redshifts, though quiescent galaxies at high redshift and high stellar masses demonstrate strong negative $r_{e, \mathrm{mass}}/r_{e, \mathrm{light}}$ ratios. The lower two panels depict the $r_{e, \mathrm{mass}}/r_{e, \mathrm{light}}$ ratios dependent on stellar mass. Though the mass bins for which the average ratios are calculated are the same for both populations, the data points of the young quiescent galaxies are offset horizontally to improve on readability.}
    \label{fig:rad_mass}
\end{figure*}

\citet{clausen20243ddash} show that the evolution of the half-light radii is mass-dependent, such that the most prominent difference between the young and old quiescent populations is only evident for the most massive galaxies in the sample with $M_* >10^{11} M_{\odot}$.
Due to the small volume of the PRIMER survey, the sample we consider is particularly devoid of massive galaxies. 
In fact, the sample includes only 9 young quenched galaxies above this threshold. 
To gain insight into the effects of total stellar mass in Figure \ref{fig:rad_mass}, we show the dependence of both half-light and half-mass radius (top panels), as well as the resulting color gradients (bottom panels) on the total stellar mass in three mass bins.
To mitigate the effect of probing different stellar populations at different wavelengths we use rest-frame 0.7$\mu$m half-light radii.
The left panel shows the measurements of the average half-light (triangles) and half-mass (circles) radius at 0.5$<$ z $<$ 1.75, whereas the right panel shows the same for galaxies at 1.75 $<$ z $<$ 3.
As expected, both half-light and half-mass radii of both young (blue) and old (red) quiescent galaxies increases with stellar mass, congruent with previous measurements of the quiescent galaxy population \citep[e.g.,][]{Suess2020,mowla2019, mosleh2020, clausen20243ddash, Martorano2024}.

The bottom panels display the average ratio of $r_{e, \mathrm{mass}}/r_{e, \mathrm{light}}$ for each bin.
The error bars display the error in the mean with small errors in the young quiescent population caused typically by a coincidental grouping of measurements.
Whereas small error bars in the old quiescent galaxies are connected to the larger numbers of galaxies in the respective bins.
In agreement with the general evolution observed in section \ref{subsec:half-mass_radii_evolution}, both half-light and half-mass radii are, on average, larger at lower redshift than at high redshift for both old and young quiescent galaxies at all masses considered herein.

While the data show no trends with mass in the color gradients for old quiescent galaxies (spearman $\rho = -.08$, $p$-value$= 0.09$, young quiescent galaxies exhibit a slight trend towards increasingly negative color gradients with increasing stellar mass at $0.5<z<1.75$ for young quiescent galaxies (spearman $\rho = -.58$, $p$-value$= 0.01$).
However, the presence of this trend is heavily dependent on the number of mass bins and their boundaries chosen to calculate the color gradients. 
The ratio of $r_{e, \mathrm{mass}}/r_{e, \mathrm{light}}$ for both young and old quiescent populations seems to stay relatively constant with stellar mass and redshift for the range of $1.75<z<3$ (spearman $\rho = -.24$, $p$-value$= 0.27$ for young quiescent and (spearman $\rho = -.34$, $p$-value$= 0.01$ for old quiescent galaxies).
The scatter combination with the low number of massive, young quiescent galaxies suggests that the two populations to agree in color gradients at all masses and redshifts explored within the uncertainties.

Previous studies have interpreted small $r_{e, \mathrm{mass}}/r_{e, \mathrm{light}}$ ratios as evidence of ''outshining'': strongly negative color gradients indicate inhomogeneous stellar populations with younger, brighter stars at larger radii, flattening the surface brightness profile while not significantly affecting the stellar mass profile.
However, this outshining effect should, to first order, impact only the recently quenched population and have no significant effect on the older quiescent galaxies by virtue of the respective ages of their stellar populations.
If the size difference observed in massive young and old quiescent galaxies results from outshining, the older quiescent galaxies should have neutral color gradients at these high masses. 
In contrast, the younger population should exhibit strongly negative color gradients.
According to our measurements, this does not seem to be the case, which leads to the conclusion that the observed trends in size measurements are intrinsic to the population and not impacted by ''outshining''.

\subsection{Evolution of Color Gradients}\label{subsec:half_mass_evolution_colorgradients}

The previous sections use rest-frame wavelengths to determine light-weighted size measurements. 
However, earlier studies of size evolution and color gradients are based on observed wavelengths.
To understand the impact and limitations of probing different parts of the spectrum when measuring half-light radii in a single filter for a range of redshifts, this sub-section repeats the analysis in Section \ref{subsec:colorgradient_restframe} but instead of rest-frame uses $r_{e, light}$ uses half-light radii measured at constant fixed observed wavelengths.

\begin{figure*}
    \centering
    \includegraphics[width=\textwidth]{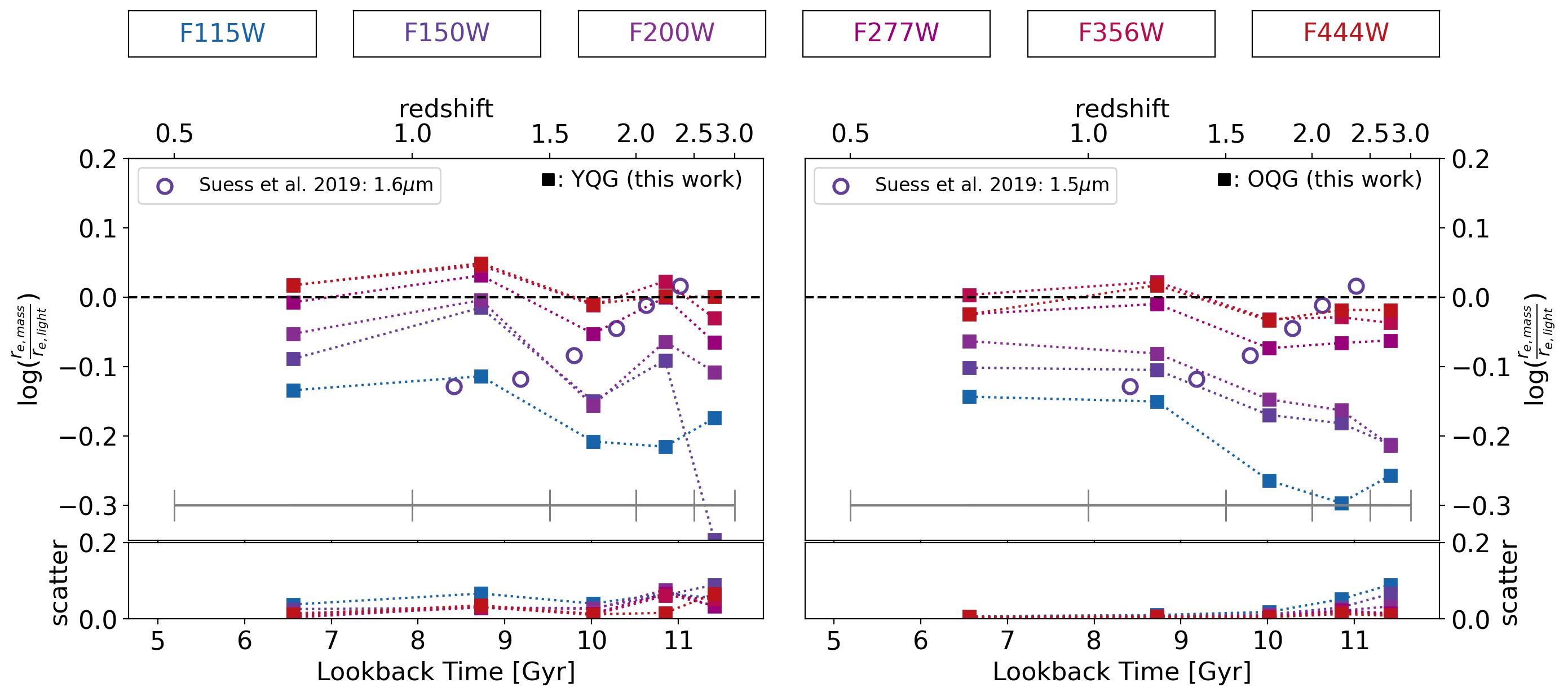}
    \caption[Dependence of the color gradients on the wavelengths used to determine the half-light radius and redshift]{Dependence of the color gradients (\textbf{upper panels}) and their scatter (\textbf{lower panels}) on the wavelengths used to determine the half-light radius and redshift. For both young quiescent galaxies (\textbf{left panels}) and the old quiescent galaxies (\textbf{right panels}), the color gradient increases with the wavelength at which the half-light radius is measured. The color gradient of the old quiescent population does not seem to evolve much with redshift. In contrast, the color gradients of young quiescent galaxies are larger at z $>$ 1.75 independent of wavelength of $r_{e, light}$. Measurements taken from \cite{suess2019a} at F160W (1.6$\mu$m) and rest-frame 0.5$\mu$m half-light radii taken from \cite{vanderwel2024} are included.}
   \label{fig:colorgradient_wavelength}
\end{figure*}

Figure \ref{fig:colorgradient_wavelength} shows the evolution of color gradient for young quiescent (left panel) and old quiescent (right panel) galaxies with the extent of the bins demonstrated by the black bar at the bottom.
The data points are color-coded by the filter bandpass where the $r_{e, light}$ measurement was made, with redder colors corresponding to measurements taken at longer wavelengths.

The data clearly show a strong dependence of color gradients on the wavelength at which half-light radius is measured for both populations.
Longer wavelengths lead to more neutral $r_{e, \mathrm{mass}}/r_{e, \mathrm{light}}$ ratios.
This trend is consistent with the previous observations demonstrating a decrease of measured half-light radii at longer wavelengths \citep[e.g.][]{Suess2022,vanderwel2024} and concurs with the observation that F444W half-light radii agree best with half-mass radii measurements for old quiescent galaxies.
This wavelength dependence does not appear to affect the general trends of evolution observed. 
For each population, the shape of the color gradient evolution appears to be largely independent of the wavelength used to determine the half-light radius.
The significantly stronger negative color gradient in the highest redshift bin measured in the F150W filter is the result of a single galaxy that displays a larger than average half-light radius in this band. 
The nature of this peculiar galaxy requires further investigation that is beyond the scope of this paper.
While it does lower the mean color gradient of young quiescent galaxies at this redshift range, it does not impact the general trends observed herein.

There appears to be little to no evolution of color gradients with redshift in the young quiescent galaxy population (spearman:  $-0.4 <\rho<-.3$ with $p$-value $\approx 0.03$, independent of filters).
At face value, the color gradients of old quiescent galaxies seem to decrease towards more negative color gradients at high redshift in F115W, F150W and F200W.
However, calculating the Spearman correlation coefficient reveals $\rho = -0.18, -0.20, -0.33$ with associated $p$-values of 1e-4, 2.57e-5, 1.7e-12, respectively. 

At first glance, this behavior differs from measurements of the total quiescent population from \cite{suess2019a} (Figure \ref{fig:colorgradient_wavelength}, circles)  who measure the color gradients for 1.6$\mu$m and rest-frame 0.5$\mu$m, respectively.
They observe a general increase in $r_{e, \mathrm{mass}}/r_{e, \mathrm{light}}$ towards earlier cosmic times.
This conundrum can be resolved by comparing with the overall number density evolution of old and young quiescent galaxies:
Old quiescent galaxies dominate the population at lower redshift, making up $>$95\% 
Towards earlier cosmic times, the percentages of young quiescent galaxies increase significantly until they make up half the population at z $>$ 2 \citep[e.g.,][]{whitaker2012,clausen20243ddash}.
\citet{gould2023} predict based on the full COSMOS2020 catalog \citep[][]{weaver2022} a further increase with young quiescent galaxies making up 65\% at $3<z<4$ and 86\% at $4<z<5$.
Therefore, they contribute significantly to the evolution of average color gradients at early cosmic times.
Comparing the data by \cite{suess2019a} with the evolution of the young quiescent population at $z>1.75$, we find that they agree within the corresponding uncertainties with the respective measurements conducted at similar wavelengths in this work.
Similarly, at lower redshifts ($z<1.75$) the contributions of the young quiescent galaxies to the general quiescent population diminishes.
Comparing the data taken from \citet{suess2019b} to our data for the old quiescent galaxies at these redshifts exhibits excellent consistency.

We, therefore, conclude that earlier observations of increasing $r_{e, \mathrm{mass}}/r_{e, \mathrm{light}}$ ratios towards earlier cosmic times are strongly impacted by the increasing contribution of young quiescent galaxies.
However, due to the small sample of this work, a conclusive test of whether part of the trend to more neutral color gradients at larger redshifts in intrinsic to the population itself requires a larger sample of young quiescent galaxies than the one at hand.
\section{Discussion}\label{sec:half-mass-discussion}
In this paper, we compare the evolution of half-light radii, half-mass radii, color gradients, and their respective dependence on total stellar mass for a sample of recently quenched (young) and older quiescent galaxy populations.

We find that both populations significantly grow in half-light and half-mass radii by 0.3 dex towards lower redshift, with no significant difference between old and young quiescent galaxies.
The low stellar mass limit (log$M_*/M_{\odot}>10$) combined with the small covered area causes the sample to be dominated by intermediate-mass galaxies.
As such, the observation of no significant difference between young and old quiescent galaxies agrees with the result presented in \cite{clausen20243ddash}.

We also observe that the $r_{e, \mathrm{mass}}/r_{e, \mathrm{light}}$ ratios measured at observed-frame wavelengths of old and young quiescent galaxies quiescent galaxies are, on average, constant over cosmic time.
If younger stars at large radii in the recently quenched population were the reason for the observed difference in half-light radii, as suggested by \citet{suess2019a}, we would expect stronger negative color gradients in this population than in the old quiescent galaxy population.
However, this does not seem to be the case for our sample, leading to the conclusion that any size difference or lack thereof observed in light-weighted or mass-weighted measurements is intrinsic to the population and not an artifact of the chosen methodology.

The observed $r_{e, \mathrm{mass}}/r_{e, \mathrm{light}}$ ratios also demonstrate that half-light radii measured at observed 4.4$\mu$m follow the half-mass radii closely.
Since measurements at infrared wavelengths trace less massive and longer-lived stars which make up most of the mass of a quiescent galaxy, \cite{Suess2022} use rest-frame infrared sizes measured in this filter as a proxy for mass-weighted sizes.
In this paper, we undertake the extra step to directly compare 4.4$\mu$m sizes to half-mass sizes and confirm that this method is valid for both young and old quiescent galaxies across the redshift range examined herein.

If the growth of the red sequence could be solely attributed to mass accumulation through minor merger events, the old quiescent galaxy population should be larger in half-mass radii compared to the more recently quenched population.
In contrast, progenitor bias would predict the half-light radii of old quiescent galaxies to be smaller than those of young quiescent galaxies, thus driving the size evolution of the population as a whole over time.
The observed agreement of both populations in mass-weighted and light-weighted size measurements both in the rest-frame and observed-frame strongly points towards a combination of the two mechanisms driving the growth of the red sequence at intermediate masses:
At lower redshift, star-forming galaxies are larger than the ones at earlier cosmic time, naturally propagating the size growth to the young quiescent galaxy population after quenching.
At the same time, minor mergers cause the growth in old quiescent galaxies thereby closing the gap in sizes between the two populations predicted by progenitor bias.

The multi-wavelength NIRCam imaging in the PRIMER survey allows the derivation of mass profiles for a large quiescent sample of 455 galaxies. 
Unfortunately, it lacks significant coverage to detect a sizeable sample that could meaningfully represent the most massive young quiescent galaxies (log$M_*/M_{\odot}>11$).
Only 9 out of 455 quiescent galaxies fit this criterion, allowing intermediate-mass galaxies to dominate this sample and possibly cause the lack of a deviation in half-mass radii.
However, applying the method presented herein to wider area JWST surveys like the  COSMOS-Web survey \citep[][]{Casey2024} would provide a more comprehensive sample and possibly allow further insights into the matter.

Considering the scatter, section \ref{subsec:half_mass_mass_dependence} suggests that old and young quiescent galaxies have similar color gradients at all masses considered.
If true, an observed size difference in the rest-frame optical color would be intrinsic to the population and not an artifact of light-weighted size measurements.
\cite{Suess2020} demonstrate that the difference in size between quiescent galaxies with stellar ages less than 1 Gyr and those with more than 1 Gyr (same definition for old/young quiescent galaxies as our study) is less dominant if galaxies with log$M_* > 11M_{\odot}$ are not present in this sample.
They also find old quiescent galaxies with more negative color gradients than their younger counterparts, indicating that the smaller observed sizes of young quiescent galaxies are due to "outshining".
They note that their old quiescent galaxies are also more massive, which could bias their sample's age-color gradient relation towards steeper negative color gradients.
However, their sample of recently quenched galaxies that satisfy the mass limit for the most massive bin is similarly limited as our study.
A larger sample than the one at hand is needed to warrant statistical validity and properly test whether old quiescent galaxies have more negative color gradients and are, therefore, not intrinsically larger than their younger counterparts.

Another potential systemic uncertainty in our study lies in the assumption of a color-$M_*/L$ relation and the hypothesis that such a relation derived from integrated colors and $M_*/L$ ratios would also trace radial color gradients within a galaxy.
While the color-$M_*/L$ relation method to convert SBPs to mass profiles has been widely used in the past \citep[e.g.,][]{fang2013,suess2019a,suess2019b,mosleh2020,miller2023,vanderwel2024}, all of these studies agree that there is significant scatter in $M_*/L$ at fixed color.
\citep[e.g.,][]{vanderwel2024} cite a scatter of 0.12 dex, which is of similar magnitude as the uncertainty in SED-based $M_*$ estimates.
By utilizing rest-frame NIR observations that have been shown to trace mass profiles better \citep[][]{Suess2022} and carefully choosing the two available filters with the least amount of scatter around a color-$M_*/L$ relation and then fitting a redshift dependent relation, we can reduce the scatter to 0.057 dex. 
We do not find a remaining dependency of this scatter on redshift, total stellar mass, stellar age, or flux (see Figure \ref{fig:color_ML_277_444} in the Appendix).
Nevertheless, \cite{bernardi2023} point out that the color-$M_*/L$ relation heavily depends on assumed initial mass functions (IMF) gradients and can, if included, reduce half-mass sizes by up to 0.3 dex.
Correctly accounting for different IMFs requires both spatially resolved SED fitting of each galaxy as well as a firm understanding of the IMF, which is a widely discussed topic itself.
Both of these are beyond the scope of this work but should be considered for all trends presented herein.

\section{Conclusions} \label{sec:half-mass_conclusion}
In this work, we compare the structural evolution of massive (log($M_*/M_{\odot}$)$>10$) young quiescent galaxies to their older counterparts between $0.5<z<3$ selected from the PRIMER survey focusing on the derived half-light radii, half-mass radii, color gradients and dependence on total stellar mass.
With the advantage of JWST/NIRCam imaging, we are able to determine half-mass radii for quiescent galaxies and find the following:

\begin{itemize}
    \item The previously observed growth of the red sequence is intrinsic to the population, with average half-mass radii increasing by 0.45 dex for the young and by 0.40 dex for the old quiescent galaxy populations from $z=3$ to $z=0.5$. Average half-light radii increase by 0.31 dex for young and 0.32 dex for old quiescent galaxies over the same redshift range. 
    \item Rest-frame infrared sizes agree well with the mass-weighted sizes at all redshifts considered, for both old and young quiescent galaxies.
    \item At intermediate masses (10$<$log($M_*/M_{\odot}$)$>11$), old and young quiescent galaxies show the same evolutionary trends in mass-weighted size. To test for the presence of a difference in massive galaxies (log($M_*/M_{\odot}$)$>11$), a larger sample is needed to draw secure conclusions.
    \item The effect of redshift both on the varying contribution from young quiescent galaxies to the general quiescent population as well as probing different stellar populations at different cosmic times can mimic evolution in observed-frame $r_{e, \mathrm{mass}}/r_{e, \mathrm{light}}$ ratios. When correcting for these effects, rest-frame $r_{e, \mathrm{mass}}/r_{e, \mathrm{light}}$ ratios agree between populations within their uncertainties at all redshifts explored, producing slightly more negative $r_{e, \mathrm{mass}}/r_{e, \mathrm{light}}$ ratios at lower redshift.
\end{itemize}

We speculate that the previously observed size difference in massive (log$M_*/M_{\odot}>11$) young and old quiescent galaxies are caused by the strong color gradients.
However, our limited data do not demonstrate old quiescent galaxies to have the higher $r_{e, \mathrm{mass}}/r_{e, \mathrm{light}}$ ratios at these masses compared to young quiescent galaxies to close the gap in light-weighted size measurements at high redshift and mass.
Therefore, this study emphasizes the need for large datasets with significant sampling of these rare galaxies to fully understand the nature of the size difference. 
Unfortunately, such a data set is not yet available. 
Still, it may be subsidized by combining multiple surveys, such as CEERS, COSMOS-Web, or JADES, with PRIMER to conclusively determine the effect of stellar mass on the observed size difference.
\section*{Acknowledgements}
For the purpose of open access, the author has applied a Creative Commons Attribution (CC-BY) license to any Author Accepted Manuscript version arising from this submission.
The specific observations analyzed can be accessed via \dataset[doi:10.17909/ee2f-st77]{https://archive.stsci.edu/doi/resolve/resolve.html?doi=10.17909/ee2f-st77}. 
KEW gratefully acknowledges funding from HST-GO-16259.  
The Cosmic Dawn Center is funded by the Danish National Research Foundation (DNRF) under grant \#140.
\appendix
\section{Testing the Derivation of Half-Mass Radius Methodology}
\begin{figure}[h]
    \centering
    \includegraphics[width=.6\textwidth]{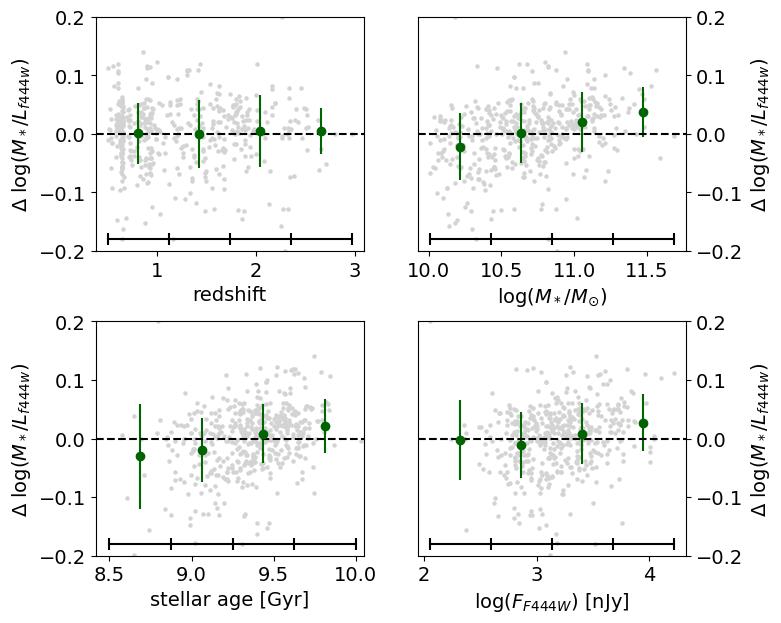}
    \caption{Depency of the derived color-$M_*/L$ relation on redshift, mass, stellar age, and total flux. There is no dependence of the relation on redshift or flux. There is a small residual dependence of the relation on both mass and stellar age, with higher masses and ages underpredicting the $M_*/L$ ratio.}
    \label{fig:testML}
\end{figure}

One of the most impactful uncertainties in this derivation method is the color-$M_*/L$ relation.
To test the accuracy of the relation, Figure \ref{fig:testML} displays the difference between the measured $M_*/L$-ratios and those predicted by the derived color-$M_*/L$ relation dependent on redshift, stellar mass, stellar age, and total flux ($log[M_*/L]_{measured}-log[M_*/L]_{fit}$).
For each aspect, we calculate the mean and standard deviation for four bins. 
We do not find a residual difference in redshift or flux.
For both mass and stellar age, there is a slight tendency of the derived color-$M_*/L$ to underestimate the $M_*/L$ ratios at small masses (younger ages) and overestimate the $M_*/L$ ratios at higher masses (older ages).

The slightly overpredicted $M_*/L$-ratios in galaxies with younger stellar ages (young quiescent galaxies) could, in theory, be responsible for the larger $r_{e, \mathrm{mass}}/r_{e, \mathrm{light}}$ ratios compared to those of the older stellar population. 
To thoroughly test this, a detailed analysis of this deviation of the fit $M_*/L$ ratios from the measured ones dependent on mass, redshift, and stellar age is needed.
Due to limited coverage, the data set at hand does not allow representative statistics for such an investigation. 
Instead, we note that though this is a caveat, the deviation at younger stellar ages is much smaller than the deviation in $r_{e, \mathrm{mass}}/r_{e, \mathrm{light}}$ ratios between old and young quiescent galaxies (0.03 dex compared to 0.1 dex).
In combination with these slight tendencies still being within the 1$\sigma$, and smaller than the uncertainty of SED-derived $M_*$ estimates, we consider the derived relation accurate enough not to impact the trends presented herein.

\begin{figure}
    \centering
    \includegraphics[width=.5\textwidth]{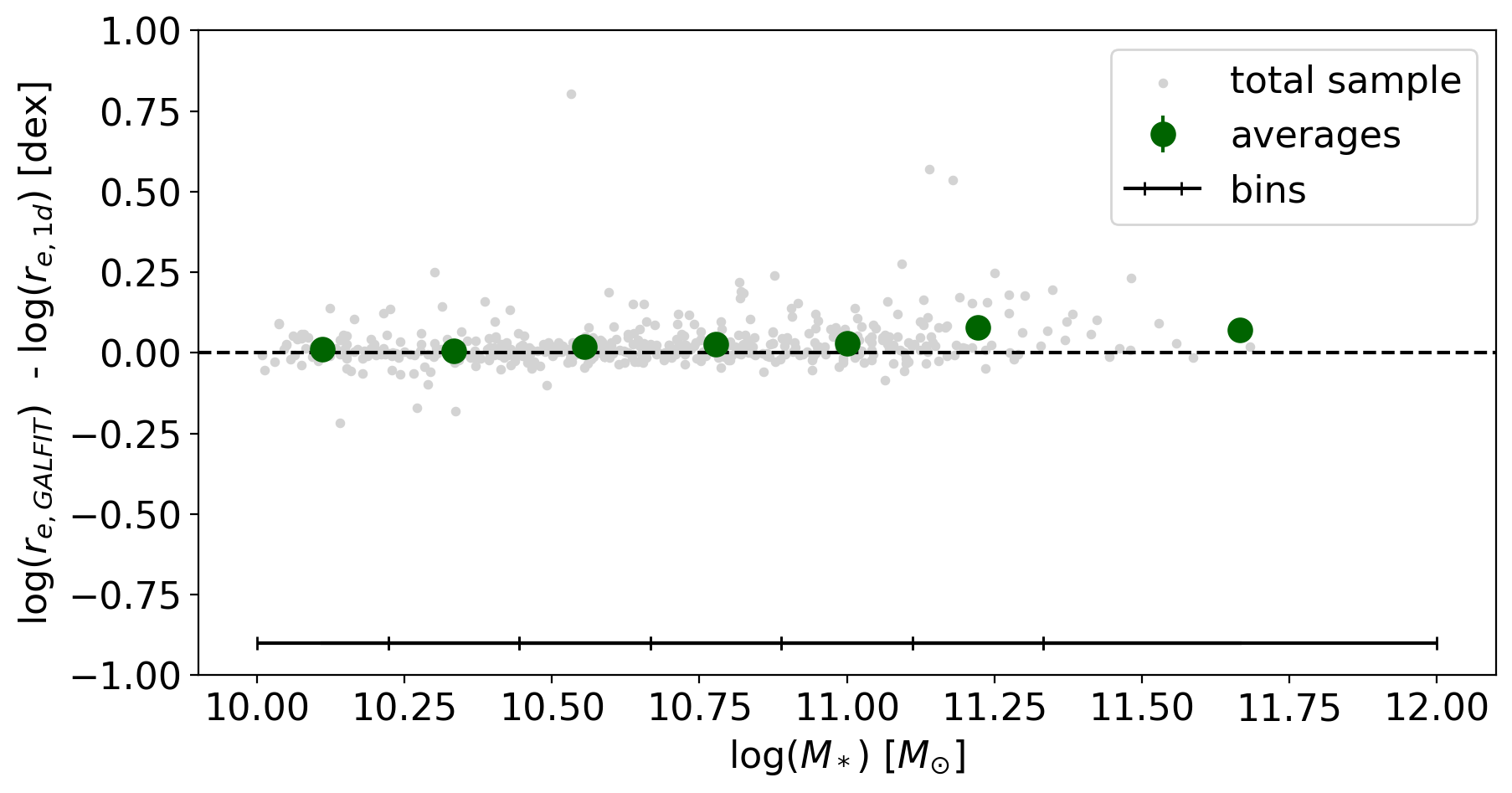}
    \caption[Comparisson of half-light radii determined with \texttt{GALFIT} and a direct one-dimensional method]{On average, there is no difference between effective radii derived from fitting a two-dimensional $Sersic$ profile to half-light radii extracted by integrating the radial SBPs. At high masses, there are small outliers where the \texttt{GALFIT} results in larger galaxies. To determine which of these methods is more accurate at large masses is beyond the scope of this work.}
    \label{fig:1d_vs_galfit}
\end{figure}

Similar to \cite{Mosleh2017}, we compare the half-light radii derived from the curve of growth of these SBP (one-dimensional) to the effective radii from the \texttt{GALFIT} fit (two-dimensional) in different bins of mass.
We find little to no significant offset (0.02dex) between the two methods at intermediate masses ($10<log(M_*/M_{\odot})<11.15$), thereby validating the one-dimensional integration method as well as the choice of $sersic$ profiles as a model (see Figure \ref{fig:1d_vs_galfit}).
At high masses (log$M_*$ $>$ 11.15$M_{\odot}$), \texttt{GALFIT} produces slightly larger half-light radii (by 0.08dex) than the one-dimensional approach which might indicate that these massive galaxies are not sufficiently well approximated by a one-component $Sersic$ profile.
If we consider the half-light radii measured in this one-dimensional approach more reliable than the \texttt{GALFIT} results, the issue of massive quiescent galaxies potentially not following a perfect $sersic$ profile is not affecting the results for the half-mass radius.
\bibliography{sample631}{}
\bibliographystyle{aasjournal}

\end{document}